\documentclass[pdftex,epj]{svjour}          % 
\usepackage{graphicx}%
\usepackage{multirow}%
\usepackage{amsmath,amssymb,amsfonts}%
\usepackage{mathrsfs}%
\usepackage[title]{appendix}%
\usepackage{xcolor}%
\usepackage{textcomp}%
\usepackage{manyfoot}%
\usepackage{booktabs}%
\usepackage{algorithm}%
\usepackage{algorithmicx}%
\usepackage{algpseudocode}%
\usepackage{listings}%
\usepackage{float}
\usepackage{cite}
\usepackage{balance}
\usepackage[colorlinks,citecolor=black,urlcolor=black,linkcolor=black]{hyperref}
\usepackage{soul, color, xcolor}
\newcommand{\CLASS}[1]{\href{https://lesgourg.github.io/class_public/class.html}{\bf CLASS}}

\newcommand{\COLOSSUS}[1]{\href{https://bitbucket.org/bdiemer/colossus/src/master/}{\bf COLOSSUS}}

\newcommand{\bibcommenthead}

\newcommand{\blue}[1]{\textcolor{blue}{#1}}

\begin{document}

\begin{sloppypar}
\begin{twocolumn}
\title{Exploring the Dark Energy Equation of State
with JWST}

\author{Pei Wang\inst{1,2} \and Bing-Yu Su\inst{1,2} \and Lei Zu \inst{1,2,}\thanks{\emph{e-mail:} zulei@pmo.ac.cn (corresponding author)} \and Yupeng Yang
\inst{3,4,}\thanks{\emph{e-mail:} ypyang@qfnu.edu.cn (corresponding author)} \and Lei Feng
\inst{1,2,4,}\thanks{\emph{e-mail:} fenglei@pmo.ac.cn (corresponding author)}
}

%%\offprints{}          % Insert a name or remove this line
%
\institute{Key Laboratory of Dark Matter and Space Astronomy, Purple Mountain Observatory, Chinese Academy of Sciences, Nanjing 210023, China \and School of Astronomy and Space Science, University of Science and Technology of China, Hefei, Anhui 230026, China \and School of Physics and Physical Engineering, Qufu Normal University, Qufu, Shandong, 273165, China \and Joint Center for Particle, Nuclear Physics and Cosmology,  Nanjing University--Purple Mountain Observatory,  Nanjing  210093, China }
\date{Received: date / Revised version: date}
%The correct dates will be entered by Springer
%
\abstract{
Observations from the James Webb Space Telescope (JWST) have unveiled several galaxies with stellar masses $M_*\gtrsim10^{10} M_\odot$ at redshifts $7.4\lesssim z\lesssim 9.1$. These remarkable findings indicate an unexpectedly high stellar mass density, which contradicts the prediction of the $\Lambda \rm CDM$ model. 
Our study utilizes the Chevallier--Polarski--Linder (CPL) parameterization, one of the dynamic dark energy models, to probe the role of dark energy on shaping galaxy formation. By considering varying star formation efficiencies within this framework, our analysis demonstrates that in a universe with a higher proportion of dark energy, more massive galaxies are formed at high redshifts, given a fixed perturbation amplitude observed today. %Furthermore, through elaborately selecting CPL parameters, the JWST observations can be successfully explained with star formation efficiencies $\epsilon\gtrsim0.05$ at a confidence level of $95\%$. 
\blue{%Furthermore, through elaborately selecting parameters without being constrained by other observations, the CPL parameterization can successfully explain the JWST observations with star formation efficiencies $\epsilon\gtrsim0.05$ at a confidence level of $3\sigma$. 
%Furthermore, when we solely  consider the JWST observations,  the CPL  parameterization can give a  successful explanation even with a low star formation efficiencies $\epsilon \sim 0.05$ at a confidence level of $3\sigma$.
%Furthermore, when we solely  consider the JWST observations,  the CPL  parameterization can give a successful explanation with star formation efficiencies $\epsilon\gtrsim0.05$ at a confidence level of $3\sigma$.
}
These intriguing results highlight the promising prospect of revealing the nature of dark energy by analyzing the high-redshift massive galaxies.
\PACS{
      {PACS-key}{discribing text of that key}   \and
      {PACS-key}{discribing text of that key}
     } % end of PACS codes
} %end of abstract
\maketitle
\section{Introduction}
\label{sec1}

James Webb Space Telescope (JWST) \cite{Gardner:2006ky} has significantly enhanced our capability to explore the space regions that were previously inaccessible to the Hubble Space Telescope (HST) \cite{freedman2001final}. This remarkable achievement provides a unique opportunity to gain deeper insights into the epoch of reionization and the formation of the first galaxies \cite{Gardner:2006ky,robertson2022galaxy,munoz2023breaking}. 
Shortly after its commissioning, JWST has successfully identified a great number of candidate galaxies at high redshifts $z \approx 6$--$16$ \cite{finkelstein2022long,naidu2022two,castellano2022early,yan2022first,donnan2023evolution,atek2023revealing,adams2023discovery,labbe2023population,rodighiero2023jwst}, indicating that galaxies in our universe began to emit photons less than 400 million years after the Big Bang. Remarkably, the initial observations from the Cosmic Evolution Early Release Science Survey (CEERS) program unveil six candidate galaxies with stunningly high stellar masses exceeding $10^{10} M_{\odot}$ at $7.4\lesssim z\lesssim 9.1$, including one galaxy with a mass of approximately $ 10^{11} M_{\odot}$ \cite{labbe2023population}. These high-redshift massive galaxies provide evidence for a star formation efficiency of $\epsilon=0.84$ at $z\approx7.5$ and $\epsilon=0.99$ at $z\approx9$ \cite{boylan2022stress}, potentially exceeding the cosmic baryon mass budget within collapsed structures.  Although further confirmation is required for these spectroscopic observations based solely on photometry, JWST has posed a challenge to the $\Lambda$CDM model \cite{haslbauer2022has,boylan2022stress,lovell2023extreme}. Various models have been proposed to address this issue %by modifying the $\Lambda$CDM model
, such as primordial black holes (PBHs) \cite{liu2022accelerating,yuan2023rapidly,su2023inflation}, dark matter \cite{gong2022fuzzy,dayal2023warm,hutsi2023did,lin2023implications}, dark energy \cite{menci2022high,wang2023jwst,klypin2021clustering}, and others   \cite{biagetti2023high,parashari2023primordial,Wang:2023xmm,jiao2023early}. Among these proposed approaches, dark energy models stand out as a potentially promising mechanism due to their simplicity and potential effectiveness.

Dark energy is introduced to drive the accelerating expansion of the universe, which is firstly supported by the observations of Type Ia supernovae (SN Ia) \cite{riess1998observational,perlmutter1999measurements}.
This mysterious component also plays a significant role in shaping the evolution of the universe, especially in the formation of structures. However, its nature remains enigmatic. The cosmological constant, first proposed by Einstein, is the most concise and natural model which associates it with the energy density of the vacuum. 
However, recent observations suggest the possibility of the dynamical evolution of dark energy rather than $\Lambda \rm{CDM}$ model \cite{zhao2017dynamical}. To portray this, various models have been proposed, including Quintessence \cite{amendola2000coupled,banerjee2021hubble}, Quintom \cite{caldwell2002phantom}, early dark energy model \cite{doran2006early,poulin2019early} and so on \cite{li2004model,Feng:2011zzo}.  Usually, we use a parametric equation to describe the evolution of the dark energy equation of state (EoS), and one of the most popular choices is the Chevallier--Polarski--Linder (CPL) parameterization \cite{chevallier2001accelerating,linder2003exploring}. This parameterization offers a flexible and model-independent framework to examine the evolution of the late universe.

Recently, Menci et al. \cite{menci2022high} have adopted the CPL parameterization to discuss the tension between the JWST observations and the $\Lambda$CDM model.
In this study, we follow the work of Menci et al. and further explore the impact of star formation efficiency on structure formation. However, our results indicate that incorporating additional dark energy within the CPL parameterization could indeed explain the presence of high-redshift massive galaxies observed by JWST, even with a low star formation efficiency $\epsilon \sim 0.05$ ($3 \sigma$ confidence level), which contradicts with the results of Menci et al.

This paper is organized as follows. In Section \ref{sec2}, we provide a brief introduction to the CPL
parameterization and the halo mass function that we adopt. We also outline the methodology for calculating the theoretical stellar mass density.
In Section \ref{sec3}, we present the results of our analysis, focusing on the impact of the CPL parameterization with varying star formation efficiencies on the process of star formation. Additionally, we also present the constraints on the parameter space of the CPL parametrization using data from JWST. Finally, in Section \ref{sec4}, we summarize our findings and draw our conclusions.

\section{Methods}
\label{sec2}

\subsection{The CPL parameterization}
\label{subsecA}
Although our knowledge on the nature of dark energy is currently limited, the EoS ($w=p/\rho$) for dark energy can still describe its property to a certain extent. %, with $p$ and $\rho$ representing its pressure and energy density respectively. 
In the CPL parameterization, $w$ is no longer a constant but varies linearly with the redshift $z$ as
\begin{equation}
    w(z)=w_0+ w_a\frac{z}{1+z},
    \label{eq:w}
\end{equation}
where $w_0$ and $w_a$ are two free parameters. The CPL parameterization will reduce to the $\Lambda$CDM model when $w_0=-1$, $w_a=0$.
%In the following work, we will fix $w_0=-1$ and adopt five different values of $w_a$ in the CPL parameterization to investigate the impact of dark energy on structure formation. 
It is evident that $w(z)$ tends to $w_0+w_a$ in the early universe ($z\to \infty$) and equates to $w_0$ at present ($z=0$).
In this parameterization, the dimensionless Hubble parameter $E(z)=H(z)/H_0$ is
\begin{align}
        E^2(z)&=\Omega_{\rm de}(1+z)^{3(1+w_0+w _a)}e^{-3w_a\left (\frac{z}{1+z}\right )} \nonumber\\
        &\quad+\Omega_{\rm m}      (1+z)^3+\Omega_{\rm r}(1+z)^4,
\end{align}
where $\Omega_{\rm de}$, $\Omega_{\rm m}$ and $\Omega_{\rm r}$ are the fractional energy density of dark energy, dark matter and radiation, respectively. 
It is
apparent that larger $w_0$ and $w_a$ correspond to more significant dark energy density and result in a larger cosmic expansion rate. All cosmological models should induce the same structure at present and the density perturbations must increase with $w_a$ at $z=7\sim 10$. Therefore, we expect to obtain high-redshift massive galaxies through increasing $w_a$.

\subsection{The halo mass function}
\label{subsecB}
Since luminous galaxies condense from gas within dark matter halos \cite{white1978core,somerville1999semi}, the choice of an appropriate halo mass function is crucial for accurate galaxy formation calculation. 
The Press--Schechter (PS) function \cite {press1974formation}, based on the spherical collapse model, has been widely used as an analytic formula for the halo mass function. However, the PS function tends to overestimate the abundance of halos near the characteristic mass and underestimate the abundance in the high-mass tail \cite{efstathiou1988gravitational,lacey1994merger,jenkins2001mass}. To overcome this limitation, various analytical halo mass functions have been proposed to fit numerical results more accurately \cite{sheth1999large,sheth2001ellipsoidal,jenkins2001mass,warren2006precision,watson2013halo,tinker2008toward,despali2016universality}. %, drawing inspiration from the original idea of the PS function.  
In this study, we adopt the Sheth--Torman (ST) halo mass function \cite{sheth1999large,sheth2001ellipsoidal}, which is based on the ellipsoidal collapse model and has been shown to  to enhance the accuracy of fitting $N$-body simulation results. The ST function is expressed as 
\begin{equation}
    f(\nu)=A\sqrt[\ ]{\frac{2a}{\pi}} \left[1+\left(\frac{1}{a\nu^2}\right )^{p}\right]\nu\exp \left(-\frac{a\nu^2}{2}\right),
    \label{eq:hmf}
\end{equation}
where the parameters $A=0.3222$, $p=0.3$ and $a=0.707$ are determined based on simulation results. In the above equation, $\nu=\delta_{\rm c}/\sigma(M,z)$, where $\delta_{\rm c}$ is the density threshold for spherical collapse, above which density fluctuations collapse and form a virialized halo. The critical overdensity $\delta_{\rm c}$ has a weak dependence on cosmological models, so we adopt a consistent value of $\delta_{\rm c}=1.686$ \cite{gunn1972infall} for various combinations of $w_0$ and $w_a$. Additionally, $\sigma(M,z)$ denotes the variance of the linear matter power spectrum $P(k)$ smoothed on the scale $R(M)=(3M/4\pi \bar{\rho})^{1/3}$ using the top-hat filter $W(x)=3[\sin(x)-x\cos(x)]/x^3$,
where $\bar{\rho}$ represents the mean density of the universe at present. Altogether, the explicit expression of $\sigma(M,z)$ is defined as
%The critical overdensity $\delta_{\rm c}$ in an Einstein-de Sitter universe\cite{gunn1972infall} can be estimated as
%\begin{equation} 
    %\delta_{\rm c, EdS}=\frac{3}{5}\left( \frac{3\pi}{2} \right)^{2/3}\simeq 1.686. 
%\end{equation}
%In flat cosmologies with dark energy, the evolution of $\delta_{\rm c}$ can be derived as %\cite{mo2010galaxy}
%\begin{equation}
    %\delta_{\rm c}(z)\simeq \delta_{\rm c,EdS} \Omega_{\rm m}(z)^{0.0055}.
%\end{equation} 
\begin{equation}
    \sigma^2 (M,z)\equiv \frac{D^2(z)}{2\pi^2} \int_{0}^{\infty} k^2P(k)W^2(kR(M))\,{\rm d}k,
\end{equation}
where $D(z)$ denotes the linear growth factor normalized to $D(0)=1$. %Since $D$ is approximately proportional to $a$, it is convenient to define a variable $G=D/a$, which is determined by the differential equation \cite{linder2003cosmic}
%\begin{equation}
    %G''+\left [\frac{7}{2}-\frac{3}{2}\frac{w(a)}{1+X(a)}\right]\frac{G'}{a} +\frac{3}{2}\frac{1-w(a)}{1+X(a)}\frac{G}{a^2}=0,   
%\end{equation}
%where $X(a)\equiv \Omega_{\rm m}(a)/\Omega_{\rm de}(a)$.
Using the ST function, we can estimate the number density of dark matter halos.  The evolution of the number density for a given mass per unit comoving volume can be described as \cite{press1974formation,bond1991excursion}:
\begin{equation}
    \frac{\mathrm{d} n}{\mathrm{d} M} = f(\nu)\frac{\bar{\rho}}{M^2} \frac{\mathrm{d}\ln \sigma ^{-1} }{\mathrm{d}\ln M}.
\end{equation}
%and the dependence on cosmology is absorbed in $\nu$.
 
To compare the number density of dark matter halos between different cosmological models, it is necessary to normalize the matter power spectrum. 
%There are mainly two normalization methods. The first one is to normalize the power spectrum to the Cosmic Microwave Background (CMB) data at redshift $z\sim1090$ \cite{mortonson2011simultaneous,wang2023jwst} with the same primordial power spectrum amplitude $A_{\rm s}$ and the spectral index $n_{\rm s}$. The secend method is normalizing the halo mass function to the same at $z=0$. These two normalization methods may  yield different results \cite{mortonson2011simultaneous}.
Two main normalization methods are commonly used.  The first one involves normalizing the power spectrum to match the Cosmic Microwave Background (CMB) data at redshift $z\sim1090$ \cite{mortonson2011simultaneous,wang2023jwst} with the same primordial power spectrum amplitude $A_{\rm s}$ and spectral index $n_{\rm s}$. The second method entails normalizing the halo mass function to be consistent at $z=0$. These two normalization approaches may produce differing outcomes \cite{mortonson2011simultaneous}.
Here, we choose to normalize the power spectrum %in the CPL parameterization 
%with various $(w_0,w_a)$ combinations 
through the second method. There are two reasons for that. Firstly, compared with the measurements of CMB anisotropy, cluster abundance measurements provide a more direct probe of density perturbation relevant to galaxy formation \cite{kuhlen2005dark}. Secondly, normalizing the primordial power spectrum makes it challenging to distinguish the impact from different combinations of $(w_0, w_a)$ on galaxy formation \cite{wang2023jwst}, as the dark energy in the CPL parameterization predominantly affects the late universe.
%through normalizing the primordial power spectrum, it is difficult to distinguish the impact from different combinations of $(w_0,w_a)$ on galaxy formation \cite{wang2023jwst}, since the dark energy in the CPL parameterization plays a significant role only in the late universe. %We can better distinguish the impact of different $(w_0,w_a)$ combinations on galaxy formation by fixing the $\sigma_8$, while variations in $(w_0,w_a)$ have little effect on stellar mass density when fixing $A_{\rm s}$ and $n_{\rm s}$ \cite{wang2023jwst}.

%by fixing $\sigma_8$. 
%measuring the linear-theory matter fluctuation amplitude today at 
%We will now explain why normalizing the halo mass function to the same value at present is roughly equivalent to fixing $\sigma_8$, as mentioned earlier.
From Eq. \eqref{eq:hmf}, it is apparent that the halo mass function is uniquely affected by the cosmological model through $\nu(z, M)=\delta_{\rm c}(z)/\sigma(M,0)D(z)$. In our analysis, we employ the following fitting formula for $\sigma(M,0)$ \cite{vianapedro1996cluster}:
\begin{equation}
\sigma(M,0)=\sigma_8 \left( \frac{M}{M_8}\right)^{-\gamma(M)/3},
\end{equation}
where $M_8$ corresponds to the mass enclosed within a sphere of radius $R_8=8h^{-1}\,\rm Mpc$, and $\sigma_8$ is
\begin{equation}
    \sigma_{\rm 8, CPL}=\frac{\delta_{{\rm c,CPL}}(0)}{\delta_{\rm c, \Lambda CDM}
(0)} \sigma_{\rm 8,\Lambda CDM}.
\end{equation}

Generally, different cosmological models with various parameters would yield different $\sigma_8$. However, since the discrepancy between $\delta_{\rm c, CPL}(0)$ and $\delta_{\rm c,\Lambda CDM}(0)$ is negligible, we can use the same $\sigma_8$ to normalize the current value of $f(\nu)$ \cite{menci2022high}. In this work, we use the PYTHON package \CLASS{1} \cite{blas2011cosmic} and \COLOSSUS{1} to calculate 
the halo mass function.

%where $\delta_{\rm c,\Lambda CDM}(z=0)$ ($\delta_{\rm c, CPL}(z=0)$) is the critical overdensity in the $\Lambda \rm CDM$ (CPL) model at present. 

\subsection{The stellar mass density}
\label{subsecC}
The stellar mass $M_\star$ contained within a particular dark matter halo is expressed by the equation $M_{\star}=f_{\rm b}M_{\rm halo}$, where $M_{\rm halo}$ is the mass of the dark matter halo, and $f_{\rm b} = \Omega_{\rm b} / \Omega_{\rm m}$ represents the cosmic baryon fraction, with $\Omega_{\rm b}$ being the fractional energy density of baryon. In order to fit JWST observations \cite{labbe2023population} within the CPL parameterization,
we introduce the maximum cumulative stellar mass density expressed as
\begin{equation}
    \rho_{\rm max}(>M_{\star})\equiv f_{\rm b} \epsilon\int_{z_1}^{z_2}\int_{M_{\rm{halo}}}^{\infty}\frac{{\mathrm{d}} n}{{\mathrm{d}} M}M\, {\rm{d}} M \frac{{\mathrm{d}} V}{{\mathrm{d}} z} \frac{\mathrm{d} z}{ V(z_1,z_2)}, 
\end{equation}
where $V$ represents the comoving volume and satisfies 
\begin{equation}
    \frac{\mathrm{d} V}{\mathrm{d} z} =\frac{4\pi d_{\rm L}^2}{H(z)(1+z)^2},
\end{equation}
with $d_{\rm L}$ being the luminosity distance.
%where $d_{\rm L}$ is the luminosity distance.

Furthermore, there is another crucial physical quantity that warrants attention: the star formation efficiency, denoted by $\epsilon$, which represents the ratio of stellar mass to baryon mass.
Previous studies ($z \approx 0$--$10$) have shown that halos with masses near $10^{12} M_\odot$ at redshift $z\sim2$ are the most efficient when forming stars, with a star formation efficiency of $\epsilon=0.2$--$0.4$ \cite{behroozi2013average,behroozi2019universemachine}. While halos with much higher or lower masses, or those at higher redshift such as JWST observation ($7.4\lesssim z\lesssim 9.1$), are supposed to have lower efficiencies\cite{behroozi2013average}. Recently, Boylan-Kolchin \cite{boylan2022stress} has found that plausible values of $\epsilon=0.1$ or $0.32$ in the $\Lambda \rm{CDM}$ model encounter challenges in explaining JWST observations.
In light of these findings, our study considers three distinct star formation efficiencies: $\epsilon=0.1$, $0.32$, and $0.5$, with $0.1$ and $0.32$ considered plausible, while $0.5$ is deemed extreme.

The cosmological parameters we use are as follows: $H_0= 67.32\,\rm km\,s^{-1}\,Mpc^{-1}$, 
$\Omega_{\rm m}=0.3158$, %(includes baryons, dark matter, and non-relativistic neutrinos),
$f_{\rm b} = 0.156$, %($\Omega_{\rm b}$ is the baryon density parameter at $z=0$),
%the slope of the primordial power spectrum, 
$n_{\rm s}= 0.96605$, 
%and the root-mean-square amplitude of the linear matter power spectrum at $z=0$ smoothed on the scale $8\, h^{-1}\,\rm Mpc$, 
$\sigma_8 = 0.8120$, $100\theta_*=1.0410$, and $\tau=0.0543$. These values are derived from the best-fit results of the Planck Collaboration for the $\Lambda$CDM model \cite{aghanim2020planck}.
%the best-fit results of Planck Collaboration in Ref. \cite{aghanim2020planck} for $\Lambda$CDM model. %the number of dark halo mainly depends on $\Omega_{\rm m}$ and $\sigma_8$, 
Due to the challenge of reconciling the JWST observations with the CMB data \cite{wang2023jwst}, we opt to use consistent parameters across different cosmological models as benchmark points instead of conducting a global parameter fitting based on CMB data.
%Since it is difficult to reconcile the JWST observations with the CMB data \cite{wang2023jwst}, we adopt the same parameters for various cosmological models as benchmark points rather than considering the global fitting of parameters using CMB data. 
This approach is an approximation method, which brings a certain degree of uncertainty.
%The discrepancy between the best-fit parameters of the CPL parameterization and those of the $\Lambda$CDM model is not evident \cite{shah2023role},
%indicating little uncertainty in our parameter selection. 
In addition, although the cosmic baryon fraction $f_{\rm b}$ may vary across different scales, it is common to adopt the average value of $f_{\rm b}=0.156$. 
%In addition, although the cosmic baryon fraction $f_{\rm b}$ is not all the same in different scales, it is common to adopt the average value of $f_{\rm b}=0.156$. %Comparing with Ref. \cite{menci2022high} with
%$f_{\rm b}=0.18$ %and $\sigma_8=0.83$, our approach induces more
%stringent constraints on the CPL parameterization. 

%In our work, we adopt the $\Lambda$CDM parameters from the Planck Collaboration 2020 \cite{aghanim2020planck} for CPL parameterization, the present-day Hubble constant for CPL parameterization, 

%\section{Normalization $\textbf{\&}$ parameters}
%\label{sec3}

\section{Results and discussions}
\label{sec3}

The theoretical maximum cumulative stellar mass density is presented in Fig.~\ref{fig: rho}, with $w_0=-1$ fixed and $w_a$ varied, alongside the JWST measurement \cite{labbe2023population} for the redshift bins of $7<z<8.5$ and $8.5<z<10$.
For a conservative star formation efficiency of $\epsilon=0.1$, neither the $\Lambda$CDM model nor the CPL parameterization with fixed $w_0=-1$ can provide a satisfactory explanation for the JWST observations, even when $w_a$ as large as $0.8$. Similarly, when considering a plausible star formation efficiency of $\epsilon=0.32$, it is also struggling to account for the JWST observations within the $\Lambda$CDM framework.
%It can be inferred that the predicted upper limit of stellar mass density in the $\Lambda$CDM model struggles to explain the JWST observations in both redshifts with a plausible  star formation efficiency of $\epsilon=0.32$. 
However, by selecting appropriate $w_a$ in the CPL parameterization, the corresponding stellar mass density can be consistent with the JWST measurement when $\epsilon=0.32$. For $\epsilon=0.5$, the stellar mass density agrees with the observations for both the $\Lambda$CDM model and the CPL parameterization, with almost all the selected values of $w_a$. Nevertheless, a star formation efficiency of $\epsilon=0.5$ (i.e. half of the gas has been converted into stars) contradicts the multi-band observations \cite{behroozi2013average,behroozi2019universemachine}. In summary, the results indicate that higher star formation efficiency $\epsilon$ and larger values of $w_a$ lead to more massive galaxies at high redshifts. 

%\blue{Altogether, it is apparent that a higher star formation efficiency $\epsilon$ induces a larger stellar mass density. . 
%Furthermore, the results also indicate that larger values of $w_a$ lead to more massive galaxies at high redshifts. (Altogether, while higher star formation efficiency can theoretically result in a larger stellar mass density, larger values of $w_a$ are more reasonable and realistic.)}

\begin{figure*}
\centering
\includegraphics[scale=0.5]{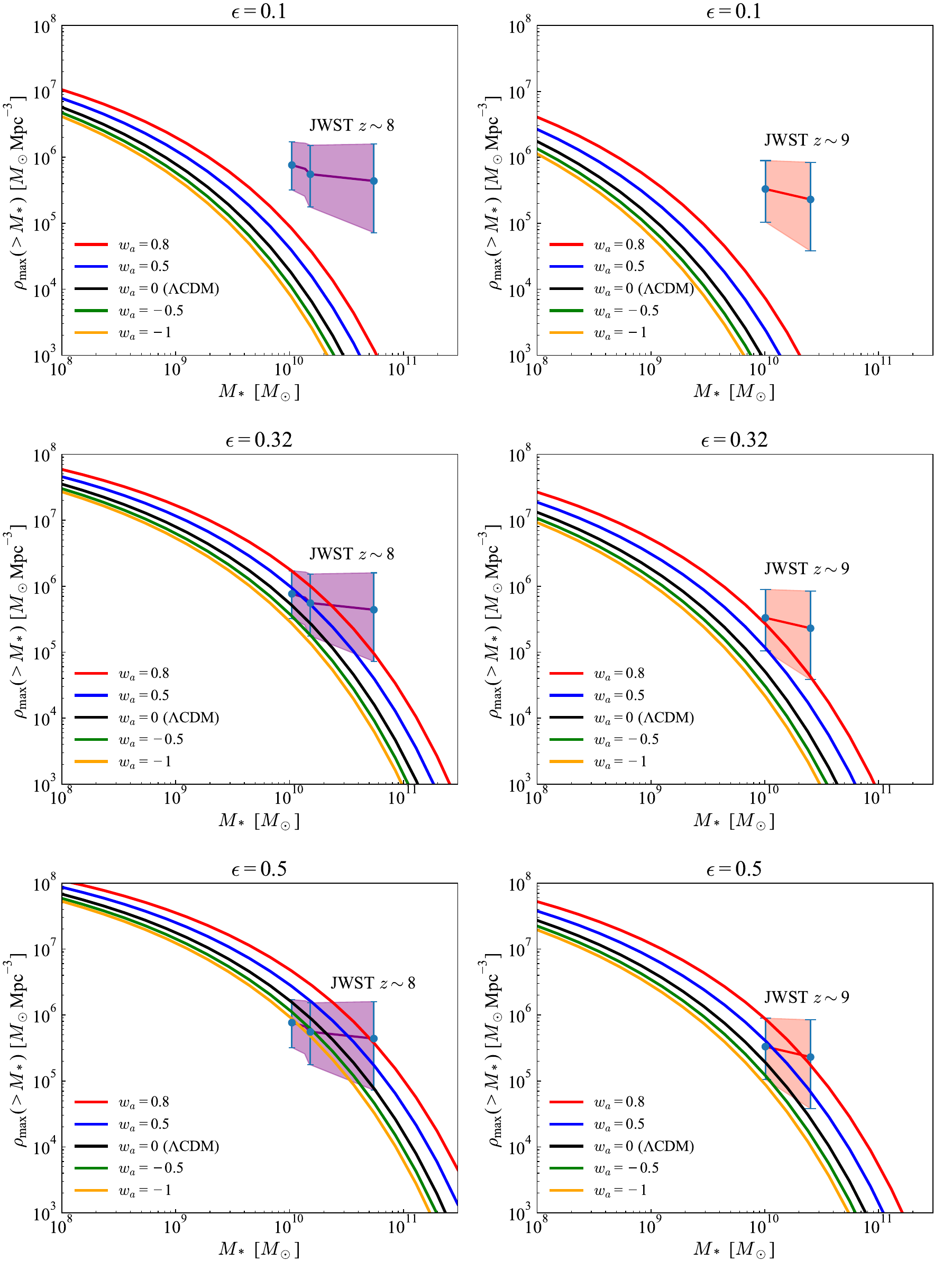}
\caption{The theoretical maximum cumulative stellar mass densities comparing with JWST observations. The CPL parameterization is applied using a fixed value of $w_0=-1$ along with five different values of $w_a$.
Solid lines depict the stellar mass density of galaxies with masses exceeding $M_*$ within two redshift bins, $7<z<8.5$ (left panel) and $8.5<z<10$ (right panel). %, corresponding to a comoving volume $V(z_1,z_2)$. 
For each redshift bin, three different values of star formation efficiency $\epsilon$ are selected for analysis. The purple and red blocks illustrate the stellar mass density within the two redshift bins, $7<z<8.5$ and $8.5<z<10$, respectively, based on the six massive galaxies observed by JWST \cite{labbe2023population}.}
\label{fig: rho}
\end{figure*} 

%Additionally, the candidates from JWST at redshift $z \sim 10$--$13$ may have stellar masses that are more massive than past predictions \cite{behroozi2020universe,harikane2023comprehensive}.

%Nevertheless, for the redshift range of $8.5<z<10$, there is still about $1 \sigma$ discrepancy between the theoretical value in the $\Lambda$CDM model and the observations, in the meanwhile, the CPL parameterization with a suitable combination of $w_0$ and $w_a$ may alleviate this tension.

To find the optimal values of $w_0$ and $w_a$ in explaining the JWST observations, we compute $\chi^2$ by scanning a two-dimensional grid ($w_a$--$ w_0$) consisting of $100\times100$ points. %The first dimension is $w_0$, spanning from -2 to -0.5, and the second dimension is $w_a$, ranging from -2.5 to 2.
The $\chi^2$ is constructed as  $\chi^2(w_0,w_a)=\textstyle \sum_{1}^{i}((\rho_{\mathrm{max}, i}(w_0,w_a)-\rho_{\mathrm{obs}, i})/\sigma_i)^2$, where $\rho_{\mathrm{max}, i}(w_0,w_a)$ is the theoretical maximum of the cumulative stellar mass density, the observed density $\rho_{\mathrm{obs}, i}$ and error bars $\sigma_i$ are obtained from Ref. \cite{labbe2023population} shown as the five blue points in Fig.~\ref{fig: rho}. After scanning the grid, we identify $\chi^2_{\mathrm{min}}$. We assume the degrees of freedom to be approximately 3, although potential correlations between the observed density  $\rho_{\mathrm{obs}}$ may result in an effective degree of freedom of less than 3. Through consulting a relevant statistical table or using software, we obtain $\chi^2_{1\sigma}$=$\chi^2_{\mathrm{min}}$+3.526 ($\chi^2_{2\sigma}$=$\chi^2_{\mathrm{min}}$+8.02)
to determine $1\sigma$ ($2\sigma$) 
confidence region. 

Under the assumption of the $\Lambda$CDM cosmology \cite{labbe2023population}, the galaxies are found by JWST within a cosmic volume by integrating over the redshift range. Hence, it is necessary to correct the observed density $\rho_{\mathrm{obs},i}$ with a volume factor $V_{\Lambda}/V_{w_0,w_a}$, where $V_{\Lambda}$ and $V_{w_0,w_a}$ are the cosmic volume in the $\Lambda$CDM model and CPL parameterization, respectively \cite{menci2022high,forconi2023early}. Similarly, the luminosity factor $d^2_{{\rm L},w_0,w_a}/d^2_{\rm L,\Lambda}$ is applied to correct $\rho_{\mathrm{obs}, i}(>M_{*,i})$ and $M_{*,i}$, where $d^2_{\rm L,\Lambda}$ and $d^2_{{\rm L},w_0,w_a}$ represent the luminosity distance in the $\Lambda$CDM model and CPL parameterization, respectively. Here $z=7.5$ and $9.1$ are picked to represent the two redshift bins mentioned above.

The favored parameter space of $w_a$--$w_0$
obtained by fitting JWST data (red regions) is presented in Fig.~\ref{fig: contour}. For comparison, the constraints from the CMB and Weak Lensing data
(green regions), and the Hubble diagram of quasars (QSO) and SN Ia (blue regions), all sourced from Ref. \cite{risaliti2019cosmological}, are also displayed in Fig.~\ref{fig: contour}.  We explore three distinct star formation efficiencies: $\epsilon=0.1, 0.32$ and $0.5$. In our work, we only consider the case of $w(z)<0$ for the reason that dark energy behaved similarly to matter or radiation in the early universe when $w(z)\geq 0$. Additionally, the scenario of $w_0<-1/(1-\Omega_{\rm m})$ within the CPL parameterization may also be considered unphysical, as it could lead to a turning point in $H(z)$, which is regarded as an exotic feature in cosmology \cite{colgain2021critique}. With a conservative value of $\epsilon=0.1$, the parameters favored by JWST data
have effectively ruled out by other cosmological observations.
%, with little overlap with the parameter space constrained by other observed cosmological data.
For $\epsilon=0.32$, the JWST observations have excluded a significant portion of the parameter space for the CPL parameterization. However, there are still small regions of parameter space consistent with other observed cosmological observations. Nevertheless, for $\epsilon=0.5$, it is noteworthy that there is a large parameter space that can simultaneously interpret JWST and other cosmological observations.

\begin{figure}
\centering
\includegraphics[scale=0.48]{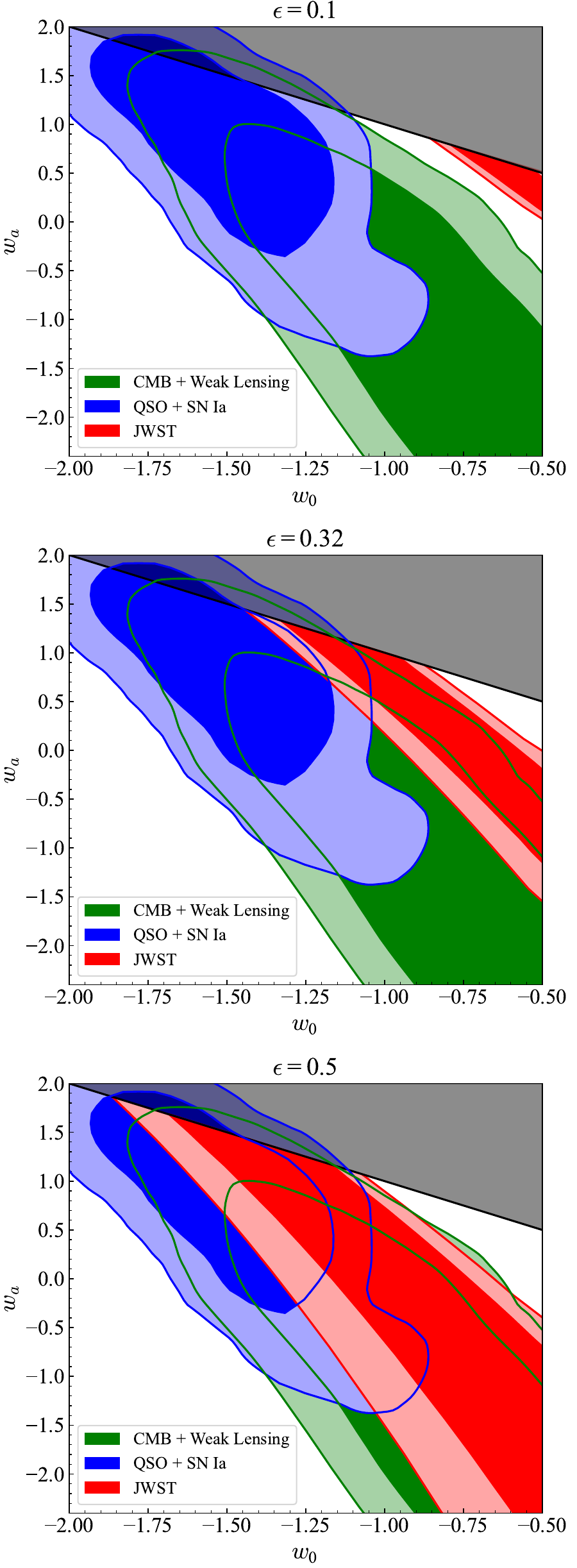}
\caption{The $w_a$--$w_0$ probability contours for three different choices of star formation efficiency $\epsilon$. The red regions represent the $1\sigma$ and $2\sigma$ confidence level by fitting the JWST data \cite{labbe2023population}, in comparison with the $2\sigma$ and $3\sigma$ contours derived from CMB and Weak Lensing data (green regions)\cite{risaliti2019cosmological}, and the Hubble diagram of QSO and SN Ia (blue regions) \cite{risaliti2019cosmological}. The black region denotes the excluded region satisfying $w(z\to \infty)\approx w_0+w_a>0$.}
\label{fig: contour}
\end{figure}

Based solely upon the JWST observations, the $\chi^2$ distribution of the star formation efficiency $\epsilon$ is presented for both the $\Lambda$CDM model and the CPL parameterization, as depicted in Fig.~\ref{fig:epsilon}.
It becomes evident that the $\Lambda$CDM model faces challenges in reconciling with the JWST observations, primarily due to its demand for an impractical star formation efficiency $\epsilon$, surpassing the upper limit of $\epsilon \approx 0.2$--$0.4$ based on previous studies \cite{behroozi2013average,behroozi2019universemachine}
. However, as shown in Fig.~\ref{fig:epsilon}, the distribution of $\chi^2$ in the CPL parameterization demenstrates a lower limit of $\epsilon=0.05$ at a $3\sigma$ cofidence level. %, suggesting that CPL parameterization can effectively account for the JWST observations even when star formation efficiency is low.%when star formation efficiencies exceed 0.05.
This finding indicates that the CPL parameterization allows for lower star formation efficiencies compared to the $\Lambda$CDM model, making it a more feasible and statistically supported model in light of the JWST data.
%Considering only the JWST observation, the $\chi^2$ distribution of star formation efficiency $\epsilon$ in the $\Lambda$CDM model and CPL parameterization are also shown as Fig.~\ref{fig:epsilon}. 
%Based on multi-band observations \cite{behroozi2013average,behroozi2019universemachine}, we focus on the star formation efficiency that lower than $0.2$, which is a conservative value. 
%Based on multi-band observations \cite{behroozi2013average,behroozi2019universemachine}, the maximum star formation efficiency is $0.2\sim0$
%Clearly, to reconcile the JWST observations, the $\Lambda$CDM model requires an unrealistic star formation efficiency $\epsilon$ which exceeds the maximum value observed in reality ($\epsilon \approx 0.2$--0.4) \cite{behroozi2013average,behroozi2019universemachine}. Whereas the CPL parameterization demonstrates its capability to explain the JWST observations with star formation efficiencies $\epsilon \gtrsim 0.05$ ($95\%$ confidence level).

\begin{figure}
%\centering
\includegraphics[scale=0.45]{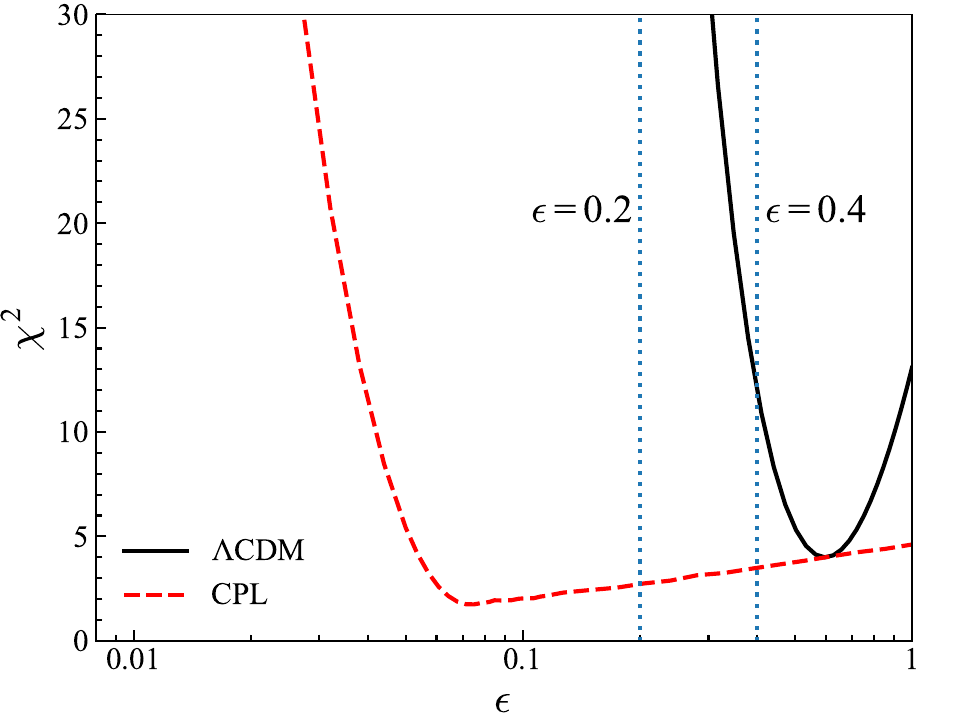}
\caption{The one-dimensional $\chi^2$ distribution of star formation efficiency $\epsilon$ in the $\Lambda$CDM model and CPL parameterization. Notably, in the case of CPL parameterization, we have adopted the minimum value of $\chi^2$ by adjusting $w_0$ and $w_a$. %through selecting appropriate $w_0$ and $w_a$ in the CPL parameterization. 
The blue dot lines represent the maximum star formation efficiency based on multi-band observations with a value of $0.2\sim0.4$ \cite{behroozi2013average,behroozi2019universemachine}.}
\label{fig:epsilon}
\end{figure} 

\begin{figure*}
\centering
\includegraphics[scale=0.5]{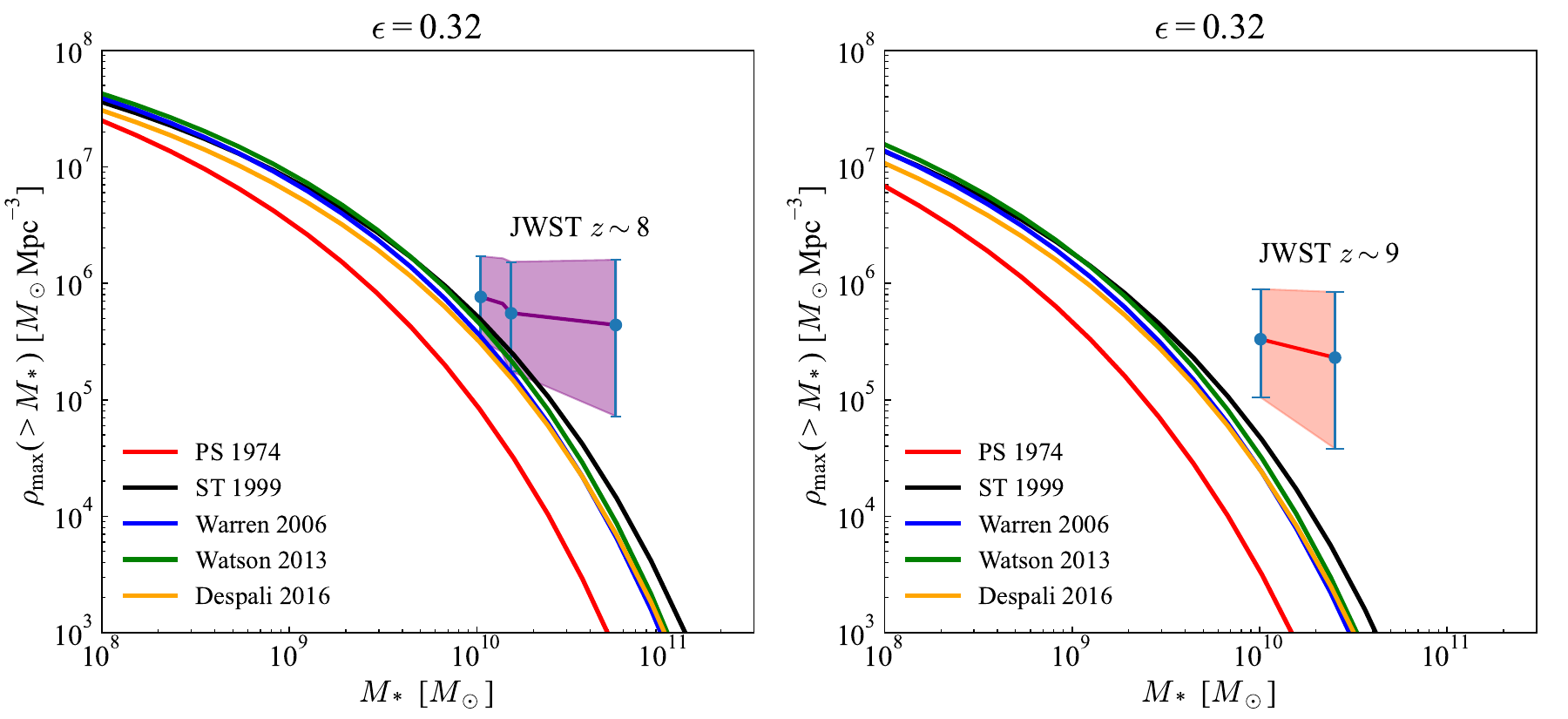}
\caption{The theoretical maximum cumulative stellar mass densities with various halo mass functions. Similar to Fig.~\ref{fig: rho}. 
However, we employed different halo mass functions, including PS 1974 \cite{press1974formation}, ST 1999 \cite{sheth1999large}, Warren 2006 \cite{warren2006precision}, Watson 2013 \cite{watson2013halo}, and Despali 2016 \cite{despali2016universality}, to examine their influence on the theoretical stellar mass densities in the $\Lambda$CDM model, with $\epsilon=0.32$ fixed.
}
\label{fig: HMF}
\end{figure*}

To assess the credibility of the ST halo mass function within the $\Lambda$CDM model framework, we explore the impact of various halo mass function models on theoretical stellar mass density. The results are illustrated in Fig.~\ref{fig: HMF}, with $\epsilon=0.32$ adopted. In this investigation, we mainly examine four well-fitting models: ST function \cite{sheth1999large}, Warren 2006 \cite{warren2006precision}, Watson 2013 \cite{watson2013halo}, and Despali 2016 \cite{despali2016universality}. The PS function \cite{press1974formation} is also included, although it diverges from the detailed results of N-body simulation. Our findings reveal that the four well-fitting models exhibit very similar behaviors, whereas the PS function deviates to some extent. Consequently, we conclude that the stellar mass density shows little sensitivity to the various well-fitting halo mass function models. Based on the above, it is considered credible to adopt the ST halo mass function as a representative.

\section{Conclusions}
\label{sec4}

The detection of high-redshift massive galaxies by JWST poses a significant challenge to the $\Lambda$CDM model.  %However, it has good agreement with current astronomical observations, including Type Ia supernovae \cite{sullivan2011snls3}, baryonic acoustic oscillations (BAO) \cite{aubourg2015cosmological}, anisotropies in the CMB \cite{komatsu2009five,ade2016planck}, and the large-scale structure of the universe \cite{reid2012clustering}.
To reconcile the $\Lambda$CDM model with JWST observations, the star formation efficiency should be extremely high \cite{menci2022high}. 
Nevertheless, when we solely consider the JWST observations, the CPL  parameterization can give a successful explanation even with a low star formation efficiencies of $\epsilon\sim0.05$. %Nevertheless, our research demonstrates that the CPL parameterization can account for the JWST observations with reasonable 
%star formation efficiency $\epsilon\gtrsim0.05$ at $3 \sigma$ confidence
%level, as long as appropriate parameters are chosen.
%Notably, assuming an efficiency of $\epsilon=0.5$, the parameter space of the CPL parameterization constrained by JWST almost overlaps with the parameter space constrained by Planck and Weak Lensing data, supernovae and quasars. %Recent studies have suggested that the luminous galaxy
%candidates at $z\sim16$ may have extremely high efficiency of $\epsilon \sim 0.6$ \cite{harikane2023comprehensive}, indicating that most of the baryons may have converted to stars at high redshift. Additionally, the candidates from JWST at redshift $z \sim 10$--$13$ may have stellar masses that are more massive than past predictions \cite{behroozi2020universe,harikane2023comprehensive}.
%However, $\epsilon=0.5$ may also be unrealistic in galaxy formation, as the star formation efficiency must satisfy $\epsilon \lesssim 0.4$ in the redshift range $z=0$--$10$ based on current observations from previous studies \cite{behroozi2013average,behroozi2019universemachine}. 
Our findings also indicate that, for a fixed power spectrum normalization (fixed $\sigma_8$), larger values of the CPL parameter $w_a$, which correspond to a greater amount of dark energy, result in higher halo mass densities and more massive galaxies at high redshifts \cite{kuhlen2005dark}. Moreover, the  constraints of CPL parameters derived from JWST are consistent with other cosmological observations for large star formation efficiency.

%The JWST data has the potential to significantly narrow down the allowed parameter space in the CPL parameterization, surpassing the constraints imposed by other cosmological probes. 
Futhermore, due to the potential tension between CMB data and JWST observations \cite{wang2023jwst,forconi2023early}, we have not incorporated CMB data in the comprehensive fitting process when interpreting the JWST observations. However, it is crucial for future research to address this issue and combine JWST observations with CMB data to achieve more precise constraints on the parameter space through global fitting.

\vspace{.5cm}

\noindent {\bf Acknowledgements}
We are grateful to Professor Xiao-Dong Li, Yi Wang, Yi-Zhong Fan, Yue-Lin Sming Tsai, Deng Wang for their helpful discussion. This work is supported by the National Key R\&D Program of China (Grants No. 2022YFF0503304), the National Natural Science Foundation of China (12220101003, 11773075), and 
the Shandong Provincial Natural Science Foundation (Grant No. ZR2021MA021).

\vspace{.5cm}

\noindent {\bf Data Availability Statement}
No Data associated in the manuscript.

%
% For  figures use
%\begin{figure*}
%\vspace*{5cm}       % Give the correct figure height in cm
%\includegraphics{leer.eps}
%\end{figure*}
% or  this
%\begin{figure}
%\centering
%\resizebox{0.75\textwidth}{!}{%
%  \includegraphics{leer.eps}
%\vspace{5cm}       % Give the correct figure height in cm
%\end{figure}
%\begin{table}
%\centering
%\begin{tabular}{lll}
%\hline\noalign{\smallskip}
%\end{tabular}
% Or use
%\vspace*{5cm}  % with the correct table height
%\end{table}

%
% BibTeX users please use
% \bibliographystyle{}
% \bibliography{}
\bibliographystyle{sn-nature}
\balance
\bibliography{ref}

\end{twocolumn}
\end{sloppypar}
\end{document}